# A WEB-BASED MULTILINGUAL INTELLIGENT TUTOR SYSTEM BASED ON JACKSON'S LEARNING STYLES PROFILER AND EXPERT SYSTEMS

H. MOVAFEGH GHADIRLI

*Young Researchers Club, Islamshahr Branch, Islamic Azad University, Islamshahr, Iran*
*hossein.movafegh@iau-saveh.ac.ir*

M. RASTGARPOUR

*Department of Computer Engineering, Saveh branch, Islamic Azad University, Saveh, Iran*
*m.rastgarpour@iau-saveh.ac.ir*

Nowadays, Intelligent Tutoring Systems (ITSs) are so regarded in order to improve education quality via new technologies in this area. One of the problems is that the language of ITSs is different from the learner's. It forces the learners to learn the system language. This paper tries to remove this necessity by using an Automatic Translator Component in system structure like Google Translate API. This system carry out a pre-test and post-test by using Expert System and Jackson Model before and after of training a concept. It constantly updates learner model to save all changes in learning process. So this paper offers an E-Learning system which is web-based, intelligent, adaptive, multilingual and remotely accessible where tutors and learners can have non-identical language. It is also applicable Every Time and Every Where (ETEW). Furthermore, it trains the concepts in the best method with any language and low cost.

## 1. Introduction

In 1982, Sleeman and Brown reviewed the state of the art in computer-aided instruction. They proposed a novel term of ITSs to describe these evolving systems as well as distinguish them from other educational systems. So the implicit assumption corresponding to the learner focused on *learning-by-doing*. They [1] classified available ITSs into four categories include of problem-solving monitors, coaches and teachers, laboratory instructors, and consultants.
The first E-Learning software was static and non-intelligent. A course had been organized by a prior procedure and taught in the same style for all learners. They were either computerized versions of textbooks (characterized as electronic page turners) or drill and practice monitors which were giving a learner some problems and comparing the learner's answers with pre-scored answers[1].

There are several kinds of Learners in the Web. It means that some learners need to repeat some lessons mean while some lessons must be removed for others. Later the researchers concluded that learning process must be dynamic and intelligent based on pedagogy view [2]. But developing an applicable and trustful system is so hard [2]. This caused to advent new generation of intelligent educational systems [3].

*Web-based learning* and *intelligent learning* is so regarded in education todays[2, 4]. A web-based tutor has some benefits like tirelessly, dominance on concepts, low cost and independent of time and place. It can utilizes conversational agents as well [3]. However millions learners of the world can learn via thousands of expert tutor through the web in an intelligent and virtual school. One of the important classes of intelligent tutors is based on rule-based expert system. It determines whatever the student knows, doesn't know and knows incorrectly. It uses this information to adjust learning style. This education style which enjoys the benefits of Expert Systems is called *model tracing tutor* [5].

ITS sometimes has different language of learner language. So such participate needs to know its language in this virtual class. On the other hand, the learning of a foreign language is often difficult and impractical for adults. Moreover, many such courses can be very time consuming, because learners often must cope with unfamiliar writing systems as well as differing cultural norms. So a system is demanded in order to eliminate this "language barrier".

This paper introduces an intelligent system to enjoy Expert Systems abilities. So E-Learning would be efficient, adaptive and just needs a computer and internet connection. Adapting with web-based contexts is very important, because the concept, which is developed for one user, isn't useful for others [2].

The proposed system determines the learning style through a test. It develops a primary model of learner. The learning process starts then. Some characteristics of learner may be varying during learning progress gradually. These progresses would be saved in learner model by system. So learner model gets closer step by step. The system can receives scientific and mental feedback of the learner and change the learning style during the process.

A learner doesn't have to know extra language by using an *Automatic Translation* component. The learner can be read the contents of E-books, virtual blackboard, chat rooms and even write with any known language.

Web-based learning also is useful for training new employees and adapting staffs according to company changes. However proposed system is an artificial environment that easily loses the motivational benefits of authentic task-oriented dialog, but ideally, web-based learning improves the learning efficiency for students and employees through features that are not available in face to face learning. It also allows learners to access the learning materials and interact with the rest of course ETEW. The aim of proposed system is to offer the content which the learner can't be aware of it easily with any language.

The rest of this paper is organized as follows. Section II defines an intelligent tutoring system, automatic translator and presents some available samples. It also deliberates the role of pedagogy in learner modeling Jackson Model. Section III describes the proposed E-Learning system which is intelligent, adaptive, customized, web-based and multilingual. Finally this paper concludes in Section IV.

## 2. Background

### 2.1. *Intelligent Tutoring System*

ITSs are computer-based instructional systems with models of educational content. They specify what to teach, and also teaching strategies that specify how to teach [6]. In late 1960, the ITSs have moved out of academic labs and have been applied in classrooms and workplaces. Some of them have shown to be high effective [5]. Unfortunately, intelligent tutors are difficult and expensive to build whereas they are more common and have proven that to be more effective. Intelligent systems can recognize the learner type, choose appropriate course content from knowledge base and present it to learners in proper style. It also attempts to simulate a human tutor expertly and intelligently. Students using these systems usually solve problems and related sub-problems within a goal space, and receive feedback when their behavior diverges from that of the learner model.

Some design factors have role for each ITS which including component expert simulator, tutor software, learner model, modeler, and knowledge base. **Error! Reference source not found.** illustrates them.

ITSs allow "mixed-initiative" tutorial interactions, where learners can ask questions and have more control over their learning. Some available ITSs are introduced in **Error! Reference source not found.**1.

Some web-based ITSs are presented in Table 2.

### 2.2. *Automatic Translation*

Automatic Translating System is more efficient than professional human translators in terms of cost and speed of translation. Two types of translation tools which are used frequently are as follows:

- ***Machine Translation*** is suitable for e-learning localization and multilingual information retrieval.
- ***Translation Memories*** takes a certain source text and then stores it with corresponding translation done by a human. It can automate the translation process completely in desirable mode.

Machine translation has been appeared in 1981 for personal computers. In 1997, *Babel Fish* appeared the first, free and translation service on the World Wide Web [7].

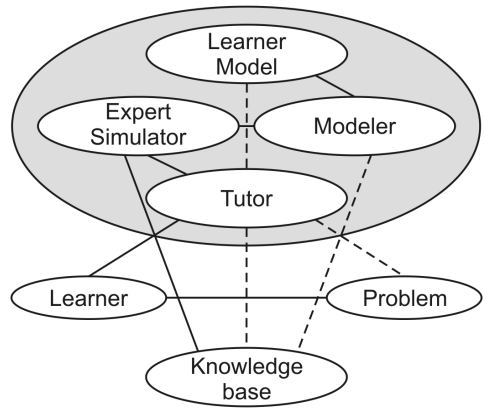

Figure 1. Intelligent Tutoring System Components

The *Automatic Translator* includes a Natural Language Parser (NLP). It can annotate phrases with structural information and refer to relevant grammatical explanations. An Error Model which detects and analyzes syntactic mistakes would be referred as well.

Table 1. Some Available Intelligent Tutoring Systems [8]

| Name | Comments |
|---|---|
| *CAES system* [9] | Has been developed by integration of shipping simulation and intelligent decision system. |
| *ICATS* [10] | Coordinates an expert system with multimedia system in an intelligent learning system. |
| *ANDES* [11] | Trains Physics without using natural language. |
| *ATLAS* [12] | Uses natural language to teach Physics. It can answer to learner's questions. |
| *ELM-ART* [13] | A web-based tutoring System for learning programming in LISP. |
| *APHID-2* [14] | A hypermedia generation system with adaption by defining rules that map learner information. |
| *AUTOTUTOR* [15] | Be full intelligent and teach *"introduction to computers"* in at least 200 universities of the world. |

The main task of the *Automatic Translator* component is real-time translating of text contents from one language to another. The teacher can store contents with any language in the system without user intervention. It will translate to the learner language then.

Table 2. Adaptive and Intelligent Technologies in Web-based ITS SYSTEMS [16]

| System | Hypertext Component | Adaptive Sequencing | Problem Solving Support | Intelligent solution Analysis | Adaptive Presentation |
|---|---|---|---|---|---|
| AST | Y | Y | | | Some |
| *InterBook* | Y | Y | Server | Y | Some |
| *PAT-InterBook* | Y | Y | | | |
| *DCG* | Y | | | | |
| *PAT* | N | | | Y | |
| *WITS* | Y | | | | Y |
| *C-Book* | Y | Some | | | |
| *Manic* | Y | Some | Server | Y | Y |
| *Proposed System* | Y | Y | | | Some |

It should be noted that multilingual meetings usually is online and also involve more than two learners with different languages. Automatic translation can be applicable in chat among the learners. Table 3 presents some applications which provide automatic translation for instant messaging (Chat).

Table 3. Online Chat Systems with Automatic Translation [17]

| Application | Languages |
|---|---|
| Amikai | 9 |
| Annochat | 4 |
| ChatTranslator | 7 |
| Free2IM Hab.la | 13 |
| Realtime Chat | 41 |
| IBM Lotus | 7 |
| Sametime | 7 |
| MeGlobe | 15 |
| WorldLingo Chat | 15 |

Any *Automatic Translator* needs to a web-based multilingual dictionary to be useful for online and real-time multilingual translation. Nowadays, the most of web services are provided by online dictionaries supporting two or more languages.

One of the most popular web based dictionary is *Google Translate* to translate each pair of 64 languages to each other such as English to Spanish, French to Russian, Chinese to English, etc. There are 4032 language pairs to translate.

*Google Translate*[*] is based upon a statistical translation system unlike other available machine translation systems. In fact a language model is trained on billions of words out of equivalent text in many different languages. For example, it compares a German word like "Bible" versus a Russian word like "Book". Then it uses comparison results efficiently [18].

The translation of comprehension is still no perfect because the accuracy of Google Translate varies with sentence, vocabulary complexity and by language.

### 2.3. *The Role of Pedagogy in Learner Modeling*

During the 1980's, computer scientists especially in AI continued to focus on the problems of natural language, learner models, and deduction.

The *pedagogical strategies* are issued from psychological and didactic research. It pertains with the underlying learning theories, i.e. behaviorism, cognitivism and constructivism. These strategies are specified by a pedagogical specialist – expert in education – and not a computer scientist [18].

The learner's skill profile is recorded in a Learner Model. Jackson models the learner's behavior based on psychology of personality. The Expert modeler monitors learner performance based on Jackson's learning styles profiler.

In fact, pedagogy is a bridge in the gap between learning and modern technologies. It caused ITS helps learner by "*the best way*".

### 2.4. *Jackson's Learning Styles Profiler*

Learning styles are various approaches or methods of the learning. An ITS is based on accurate recognition of behaviors and individual characteristics. There are some important factors in learning style such as aptitude, personality and behavior [11].

It's worth to be noted that the concept of "learning style" has emerged as a focal point of much psychological research. In our proposed learner model, we have adopted the Jackson's learning styles profiler in order to model learning styles of the learners. Jackson proposed five learning styles which are summarized in **Error! Reference source not found.**4. Reader can find more details in [2].

---

[*] http://translate.google.com/

Table 4. Summarization of Learning Styles [8]

| Learning Style | Specifications |
| --- | --- |
| Sensation Seeking (SS) | Believing that the experiences create learning. |
| Goal Oriented Achievers (GOA) | Self-confident to achieve difficult and certain target. |
| Emotionally Intelligent Achievers (EIA) | Rational and goal-oriented. |
| Conscientious Achievers (CA) | Responsible and insight creator. |
| Deep Learning Achievers (DLA) | Interested in learning highly. |

There are some approaches to model learner's behavior. For example Entistle approach [21] tries to connect the psychology concepts into the effective variables on the learner's view to *'learning'*. They are rarely applied in adaptive education systems despite some benefits in education and psychology of personality.

This section introduces Jackson model [22]. Jackson's model is based on the most recent researches in the psychology of personality. It also is an efficient analyzer. So this paper just applies it among available models.

Jackson model has been proposed in Queensland University [22]. This model has been developed by investigating personality, learning and evaluating for fifteen years. The learning is based on new neurological psychology in this model.

This model has some advantages. For example, it can export and report the individual specifications of the learner. It is obtained on basis of the points which the learner earns in the learning style (III-D).

The aim of this model is to recognize an ideal style based on pedagogy principles which can model learner's behavior and specifications in adaptive E-Learning system. A key premise of Jackson's model is that cognitive strategies redirect Sensation Seeking to predict functional behaviors [8].

## 3. Proposed approach

ITSs build a model of the individual learner's knowledge, difficulties and misconceptions during interaction with the system. This learner model can be compared with a model of the target domain. Suitable tutorial strategies can be inferred by the system. They are appropriate for the learner according to the contents of their learner model. In other words the educational interaction will be matched with specific and unique needs of each learner [23].

A web-based ITSs is an E-Learning system based on web which can be used ETEW. The first E-Learning system has been reported on 1995 [17, 24]. It is web-based and intelligent. Learning of all courses is customized well at home via web in this system. So learners can solve some examples and proper exercises ETEW. Finally he can take a course exam virtually or physically.

A large number of ITSs have been conducted in a single language so far. However, ITSs can support multiple languages with the integration of machine translation such as *Google Translate API* among 64 languages.

### 3.1. *Learning Environment*

The learning environment is explained in this section. A learner can visits website of intelligent virtual school after authentication. An intelligent Graphical User Interface (GUI) is an interface between learners and intelligent tutor. This section of system can affect learning efficiency. It should be user friendly as far as possible.

An intelligent virtual class has some benefits such as graphical properties, audio and video to make learning attractive. Moreover, some tools are available to simplify learning process. Learners can communicate well with this inanimate and non-physical system via these tools. Some facilities are Computer games, frequently asked questions (FAQ), Video chat and email.

Various learners may be logged with different language according to geographic region. Moreover all of E-learning systems can be customized for different languages. For example, the same e-learning system can be run for Iranian people with Persian contents.

This system has an *Automatic Translator* component which is able to integrate language translatability with the system. Language modeling for each learner is performed and store in the learner model. This component is developed with a MultiTranslator such Google Translate API that provides free, cheap and true translation in many languages at the same time.

Translation cost [25] and grammatical errors are some of the major limitations in localizing e-learning system. They are reduced along with growing of these systems.

This component helps learners to use the system without knowing any foreign language. Some users can also improve their skills in second language.

### 3.2. *Learning Process and Evaluation*

This system uses a three layered structure to implement a concept: *"Pre-test"*, *"Learning concept"* and *"Post-test"*.

9The pre-test includes of some questions planned by an expert tutor in advance. These questions can determine learner's primary knowledge level. It simulates learner model based on Jackson model [22]. The learning concept depends on the learner level. So the best method is determined to train a learner. A learning process starts up then. A post-test will evaluate the learner by some questions finally.

Learning evaluation is the most important factor to determine learning performance in E-Learning systems. It is performed with *Pre-test* and *Post-test* layers of the proposed system. The evaluation has two levels, *conceptual* and *objective*. The first one refers to the learner's understanding out of the lesson *concept*. The last one denotes to the learner's understanding of the lesson *topic*. The learner's knowledge level is determined by *concept level* and *objective level*.

The tutor can extract proper questions from question-base collected by an *expert system*, pre-test and post-test. He notes that a specific score is given to each question.

Question selection should satisfy some rules:
1. None of them should be repetitious even if a learner would be trained a specific concept several times.
2. The question must be planned for all sections of a concept entirely.
3. An expert tutor plans questions in all level.

The sequence, number and level of questions are determined according to learner level and learning type intelligently. Sum of scores is calculated and then learner level is determined after answering the questions. **Error! Reference source not found.**5 presents five categories of learner's knowledge level about a concept [13, 14].

Table 5. Categories of Knowledge Level [8]

| Knowledge Level | Score |
|---|---|
| *Excellent* | 86-100 |
| *Very good* | 71-85 |
| *Good* | 51-70 |
| *Average* | 31-50 |

The system modeler updates the learner's model along with questions answering. It can also save last academic status of learner and all his learning records.

The proposed system integrates *pedagogy* and *expert system* with web-based ITS structures. It develops a multilingual e-learning system by applying automatic translator algorithms which is illustrated in Figure 2. So it can be used by different nationalities. In addition of education application, it can be used as a tool to learn a second language.

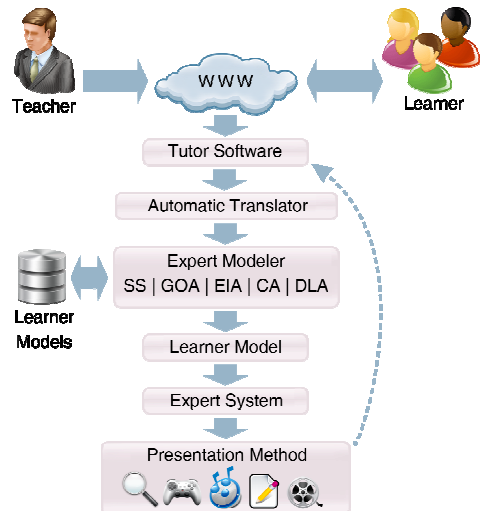

Figure 2. Block Diagram of Proposed System

## 4. Conclusion

This paper proposed a model for an adaptive, intelligent, multilingual and web-based tutor. This model can models the learning styles of the learners using Jackson's learning styles profiler and expert system technology to enhance learning performance. Previous E-Learning systems offer fixed languages and multimedia web pages and facilities for user management and communication.

Based on the proposed model, E-Learning software is implemented on the web which can identify learning styles, aptitude, characteristics and behaviors of learner in order to provide appropriate individual learning content. On the other hand, Jackson model can recognizes learner's learning style. It leads to make learner's model closer. So the efficiency of adaptation process increases.

The proposed model acts as an intelligent tutor which can perform three processes include of *pre-test*, *learning concept* and *post-test* based on learner's characteristic. This system uses expert simulator and its knowledge base as well.

Proposed model is also *multilingual.* So it helps learners to use the system without knowing any foreign language. Some users can improve their second language skills by this way.

A large number of ITS have been ever conducted in a single language. By using the proposed model, ITSs can support multiple languages with the integration of machine translation such as *Google Translate API* among 64 languages and 4032 language pairs.

The proposed system doesn't have any drawback of previous system and human expert tutor. It can improve the learning result significantly. In other words, it helps learners to study in "*the best way"* using localized language.

**References**

1. K. Eustace, "Educational value of E-Learning in conventional and complementary computing education," in Proceedings of the 16th National Advisory Committee on Computing Qualifications (NACCQ), Palmerston North, New Zealand, 2003, pp. 53–62.
2. H. M. Ghadirli and M. Rastgarpour, "A Web-based Adaptive and Intelligent Tutor by Expert Systems," in The *Second International Conference on Advances in Computing and Information Technology (ACITY 2012)*, Chennai, India, 2012.
3. P. Brusilovsky, "Intelligent tutoring systems for World-Wide Web," in *Third International WWW Conference(Posters)*, Darmstadt, 1995, pp.42-45.
4. M. a. O. Specht, R., "ACE-Adaptive Courseware Environment," *The New Review of Hypermedia and Multimedia*, vol. 4, pp. 141-161, 1998.
5. P. Brusilovsky, "Methods and techniques of adaptive hypermedia," *User Modeling and User-Adapted Interaction*, vol. 6, pp. 87-129, 1996.
6. S. Ohlsson, "Some Principles of Intelligent Tutoring," *In Lawler & Yazdani (Eds.), Artificial Intelligence and Education*, vol. 1, Ablex: Norwood, NJ, pp. 203-238, 1987.
7. J. Yang and E. Lange, "SYSTRAN on AltaVista: A user study on real-time machine translation on the Internet," *Proceedings of the 3rd Conference of the Association for Machine Translation in the Americas*, Langhorne, pp. 275-285, 1998.
8. H. M. Ghadirli and M. Rastgarpour, "A Model for an Intelligent and Adaptive Tutor based on Web by Jackson's Learning Styles Profiler and Expert Systems," Lecture Notes in Engineering and Computer Science: *Proceedings of The International MultiConference of Engineers and Computer Scientists 2012*, IMECS 2012, 14-16 March, 2012, Hong Kong, pp 63-67.
9. E. Shaw, *et al.*, "Pedagogical agents on the web," in *Proceedings Of Third International Conference on Autonomous Agents*, 1999, pp. 283-290.
10. Shute, V.J. and Regian, J.W. (1990). "Rose Garden Promises of Intelligent Tutoring Systems: Blossom or Thorn?," Presented at Space Operations, Automation and Robotics Conference, June 1990, Albuquerque, NM.

11. Koedinger, K., & Anderson, J. (1995). "Intelligent tutoring goes to the big city," *Proceedings of the International Conference on Artificial Intelligence in Education*, Jim Greer (Ed). AACE: Charlottesville, VA pp. 421-428.
12. C. Jackson, *Learning Styles and its measurement: An applied neuropsychological model of learning for business and education*: UK: PSi-Press, 2002.
13. Weber, G., & Brusilovsky, P. (2001). "ELM-ART an adaptive versatile system for web-based instruction," *International Journal of Artificial Intelligence and Education*, this volume.
14. L. Kettel, J. Thomson, and J. Greer, "Generating individualized hypermedia applications," In *Proceedings of ITS-2000 workshop on adaptive and intelligent webbased education systems*, 2000.
15. N. Entwistle, *Styles of learning and teaching: an integrated outline of educational psychology for students, teachers, and lecturers*: David Fulton Publish, 1989.
16. P. Brusilovsky, "Adaptive and intelligent technologies for web-based eduction," *KI*, vol. 13, pp. 19-25, 1999.
17. Bra PD, Santic T, Brusilovsky P. AHA! meets Interbook, and more. *Proceedings of the World Conference on E-Learning*, Phoenix, AZ, November 2003, Rossett A (ed.). AACE: Norfolk, VA, 2003; 57–64.
18. M. Aiken, *et al.*, "Automatic translation in multilingual electronic meetings, " *Translation Journal*, vol. 13, 2009.
19. M. Siadaty and F. Taghiyareh, "PALS2: Pedagogically adaptive learning system based on learning styles, " 2007, pp. 616-618.
20. C. Yang, "An Expert System for Collision Avoidance and Its Application," PH.D. Thesis, 1995.
21. J. M. Ragusa, "The Synergistic Integration of Expert Systems and Multimedia within an Intelligent Computer Aided environmental tutoring system," in *Proceedings of the 3rd world congress on expert systems*, 1960.
22. A. S. Gertner, VanLehn, K., "ANDES: A Coached Problem-Solving Environment for Physics. In Intelligent Tutoring Systems," in *Fifth International Conference*, ITS New York, 2000, pp. 133–142.
23. Y. Cui and S. Bull, "Context and learner modelling for the mobile foreign language learner," *System*, vol. 33, pp. 353-367, 2005.
24. K. F. VanLehn, R.; Jordan, P.; Murray, C.; Osan, R.Ringenberg, M.; Rosé, C. P.; Schulze, K.; Shelby, R.; Treacy, D.; Weinstein, A.; and Wintersgill,, "Fading and Deepening: The Next Steps for ANDES and Other Model-Tracing Tutors," *Intelligent Tutoring Systems*, 2000.
25. B. L. Hafsa and E. M. des Ingénieurs, "E-learning globalization in multilingual and multicultural environment," *WSEAS Proceedings NNA-FSFS-EC*, pp. 29-31, 2003.